\documentclass[showpacs,aps,pre,twocolumn]{revtex4}
\usepackage[dvips]{graphics}
\usepackage{graphicx}
\usepackage{amsfonts}
\usepackage{amssymb}
\usepackage{amsmath}

\begin{document}
\title{Dissipative Dynamics of Matter Wave Soliton in Nonlinear Optical
Lattice}
\author{F.Kh.  Abdullaev$^1$, A. Gammal$^2$,
H.L.F. da Luz$^2$ \, and
Lauro Tomio$^1$\footnote{Corresponding author
(tomio@ift.unesp.br)} } \affiliation{$^1$ Instituto de F\'\i sica  Te\'orica, 
Universidade Estadual Paulista,
01405-900, S\~ao Paulo, Brazil\\
$^2$ Instituto de F\'\i sica, Universidade de S\~ao Paulo, 05315-970, S\~ao Paulo, Brazil}
\begin{abstract}
Dynamics and  stability of solitons in two-dimensional (2D) Bose-Einstein
condensates (BEC), with low-dimensional (1D) conservative plus dissipative nonlinear optical lattices are investigated. In the case of focusing media (with
attractive atomic systems) the collapse of the wave packet is arrested by
the dissipative periodic nonlinearity. The adiabatic variation of
the background scattering length leads to metastable matter-wave solitons.
 When the atom feeding mechanism is used, a dissipative soliton can exist
in focusing 2D media with 1D periodic nonlinearity. In the defocusing media
(repulsive BEC case) with harmonic trap in one dimension and one dimensional
nonlinear optical lattice in other direction, the stable soliton
can exist. This prediction of variational approach is confirmed by the full numerical
simulation of 2D Gross-Pitaevskii equation.
\end{abstract}
\date{01/July/07}
\maketitle

\section{ Introduction}
The dynamics of optical and matter wave solitons  with different
type of management of system parameters has been under intensive
investigations in the last years~\cite{AGKT,MalomedB}. Two types
of modulations have been considered: dispersion and nonlinearity
management,  which can both occur in time and space. Temporal strong and
rapid modulations of the dispersion  are more interesting in optical
fibers due to many advantages of dispersion managed solitons for
optical communications\cite{GT,Doran,ABS} and storage of
information. Temporal modulations of nonlinearity  are promising  
in fiber ring lasers and Bose-Einstein condensates (BEC)~\cite{Berge,TM,Kartashov}. 
In the latter case, the suppression of collapse, implying in the 
existence of stable multidimensional solitons in attractive 
BEC, and the generation of periodic patterns of matter waves, 
have been predicted~\cite{AKKB,ACMK,SU,Mont,Zhar,ATMK,AG,Stefanov,adhi}. 
In optics, nonlinearity managed solitons have also been observed,
as described in Refs.~\cite{Ablow,Kevr,torres,ciattoni}.

Recently, the attention has been devoted also to the  periodic
spatial management in nonlinear optics and Bose-Einstein
condensates. In optical media, the nonlinear Kerr coefficient can be 
periodically modulated in space, leading to the problem of optical 
beam in a 2D medium with nonlinearity management. 
In BEC, the spatial variation of scattering length is
possible~\cite{AS,AGT,FK,AG05,SM,GA,BGV}, for example, by using 
optically induced Feshbach resonance~\cite{Fed,Theis}. In elongated
condensates new types of localized matter waves packets can exist.
In 2D case, the situation is less clear. The study of one-dimensional
(1D) nonlinear periodic potential in two-dimensional (2D)
non-linear Schr\"odinger equation (NLSE) shows that  broad
solitons are unstable. As verified in Ref.~\cite{Fibich1}, narrow 
solitons centered on the maximum of the lattice potential can be 
stable, but the stability region is so narrow that they are physically 
unstable. Stable gap solitons can exist in the BEC under combination 
of a linear and nonlinear periodic potentials \cite{bludov,AAG,DH}.

However, models considered till now  are strongly idealized. In
particular, using the optically induced Feshbach resonances we can 
generate mixture of conservative and dissipative nonlinear optical
lattice.  In view of that, around the Feshbach resonance, one can observed 
non-vanishing contributions to the imaginary part of the scattering length.

In this work, after an analysis of a conservative system with nonlinear 
optical lattice, we will consider  the influence of nonlinear dissipation
on the dynamics and the stability of solitons. 
In particular, we note that the role of such kind of dissipation can be 
crucial for the existence of solitons in multi dimensional nonlinear 
optical lattices. Such hope is supported by the  well known fact that 
the homogeneous nonlinear dissipation can arrest collapse in the cubic focusing
multi-dimensional NLSE~\cite{Fibich2}. The possibility of existence
of  dissipative solitons is investigated, considering 
compression effects and atom feeding.

The organization of the paper is as follows. The model is
described in the next section. In Sec. 3, it is investigated the
properties of localized states in case of attractive and repulsive 2D 
condensates in 1D nonlinear optical lattice,  with or without harmonic 
trap in one of the dimensions. In Sec. 4, it is performed an analysis 
of the evolution of 2D soliton under 1D periodic
nonlinearity and dissipation, using the variational approach and
by direct numerical simulation of the GP equation.

\section{The model}
Recently, it has been suggested to generate nonlinear optical lattices in
BEC by two counter propagating laser beams near 
 the optical induced Feshbach resonance~\cite{AG05,SM}. The spatial variation of the
optical intensity leads to a spatial periodic  variation of the
atomic scattering length. Such structure can support new types of
localized nonlinear states. The GP equation for the
wave function $\psi\equiv \psi(x_1,x_2,t)$ has the form
\begin{equation}
{\rm i}\hbar\frac{\partial{\psi}}{\partial t} =
-\frac{\hbar^2}{2m}\left(\frac{\partial^2{\psi}}{\partial x_1^2} +\beta \frac{\partial^2{\psi}}{\partial x_2^2}\right)
- g(x_1,x_2)|\psi|^2 \psi ,\label{gp1}
\end{equation}
where
\begin{equation}
g(x_1,x_2)\equiv \frac{g_{0}}{2} + (g_{1}+{\rm i} g_{2})\left(\cos^{2}(kx_1)+\delta_0\cos^{2}(kx_2)\right)\label{g}.
\end{equation}
$g_0$ is related to the $s-$wave two-body scattering length $a_s$, 
$g_0\equiv (4\pi\hbar^2/m) a_s$, with $g_0 >0$ ($g_0<0$) for attractive (repulsive) 
condensates; $g_1$ ($>0$) is related to the optical intensity; and
$g_2$ parametrize dissipative effects.

The optically induced scattering length and the inelastic
collision rate coefficient $K_{inel}$(imaginary part of $a_s$) are
described by~\cite{Fed}
\begin{eqnarray}
\Re(a_s) = a_{s0} +
\frac{1}{2k_i}\left[\frac{\Gamma_{stim}(x)\Delta}{\Delta^2 +
(\Gamma_{spon}/2)^2}\right],\\
K_{inel} \equiv \Im(a_s) = \frac{2\pi\hbar}{m}\frac{1}{k_i}
\left[\frac{\Gamma_{stim}(x)\Gamma_{spon}}{\Delta^2 +
(\Gamma_{spon}/2)^2}\right],
\end{eqnarray}
where $a_{s0}$ is the scattering length without light,
$\Delta$ is the detuning from the photo-associated resonance, and 
$\hbar k_i$ is the relative momentum of the collision.
$\Gamma_{stim}$ is the resonant transition rate between the
continuum state and the molecular state, proportional to the laser
intensity $I(x)$. Far from the resonance, the imaginary part of the
scattering length is small, such that $\Im(a) \ll \Re(a)$. 
It was shown in Ref.~\cite{Theis} that,  
in the experiment with $^{87}$Rb one can obtain optically induced 
large variations of the scattering length. The laser intensity was 
460 W/cm$^2$ and the variations $a_s$ occurred 
from $10a_0$ to $190a_0$ 
 (with $a_{s0}=100 a_0$ and $a_0$ the Bohr radius).

By considering the following variable changes and definitions in (\ref{gp1}),
\begin{eqnarray}
&&\kappa x = 2kx_1,\;\; \kappa y=2kx_2,\;\; \tau = \frac{4 w_R t}{\kappa^2},\;\;
\label{transf1} \\
&&w_R = \frac{E_R}{\hbar},\;\;
E_R = \frac{\hbar^2 k^2}{2m},\;\;
\gamma_{i=0,1,2} = \frac{g_{i}}{2|g_{0}|},\label{transf2}
\end{eqnarray}
we obtain the dimensionless equation
\begin{equation}\label{gpe0}
{\rm i}\frac{\partial{u}}{\partial\tau} +
\frac{\partial^2{u}}{\partial x^2} + \beta \frac{\partial^2{u}}{\partial y^2}
+ \gamma(x,y)|u|^2 u = 0 ,
\end{equation}
where
\begin{equation}
\gamma(x,y)\equiv \gamma_{0} + (\gamma_{1} + i\gamma_{2})(1+\delta_0
+ \cos(\kappa x) + \delta_0 \cos(\kappa y))
\label{gamma0},
\end{equation}
and the wave-function was redefined such that
\begin{equation}
u\equiv u(x,y,\tau)= \sqrt{\frac{\kappa^2|g_0|}{4E_R}}\psi.
\label{wfu}\end{equation}
From (\ref{transf2}) to (\ref{wfu}), we should note that $\gamma_0$ is
fixed to $-1/2$ for attractive condensates; and $+1/2$ for repulsive
condensates.
 Different cases can be considered:
1D geometry is realized  when $\beta = \delta_0 =0$. Anisotropic
2D case is realized for $\beta=1$, $\delta_0 =0$. And the 2D isotropic
case can be achieved with $\beta = \delta_0 = 1$.
Next, we consider more explicitly in our study the anisotropic
2D case, with  $\beta=1$ and $\delta_0 =0$.
As the soliton is completely free in
the $y-$direction, we also examine the possibility to have a harmonic trap
$m\omega_2^2 x_2^2/2$. Following the transformations (\ref{transf1}), 
a dimensionless frequency $\omega$ is also defined, such that
\begin{equation}
\omega\equiv \kappa^2\frac{\omega_2}{8 w_R},\;\;\;
\frac{m}{2}\omega_2^2 x_2^2 = \left(\frac{4E_R}{\kappa^2}\right)\omega^2 y^2.
\label{trapy}
\end{equation}
In order to extend our study of the stability conditions in a few
realistic cases, it is also verified the effect of a compression,
which can be achieved by an adiabatic time variation of the background
value of a scattering length~\cite{AS}, by modifying $\gamma_{0}$
as
\begin{equation}
\gamma_0 \to \gamma_0(\tau) = \gamma_{0}
\exp{\left[2\alpha \left(\tau-\tau_c\right)\right]\theta(\tau-\tau_c)}.
\label{gt}\end{equation}
Compression effect can also be achieved by a feeding process, 
which can be described by an additional
term $i\alpha_f u$ in  the GP equation~\cite{DK}.
Note: If the modulation of nonlinearity in time is induced by
increasing the transverse frequency of the trap, then we should
multiply the full nonlinear term by $\exp(2\alpha \tau).$
With these considerations, the Eq.~(\ref{gpe0}) can be written as
\begin{equation}
{\rm i}\frac{\partial{u}}{\partial\tau} =
-\frac{\partial^2{u}}{\partial x^2} - \frac{\partial^2{u}}{\partial y^2}
- \gamma(x,y,\tau)|u|^2 u + \omega^2 y^2 u +{\rm i}\alpha_f u,
\label{gpe1}\end{equation}
where
\begin{equation}
\gamma(x,y,\tau)\equiv \gamma_{0}(\tau) + (\gamma_{1} + {\rm i}\gamma_{2})
\left[1 + \cos(\kappa x)\right]
\label{gamma}. \end{equation}
 In the above, one should take $\alpha_f=0$ when $\alpha \ne 0$ in (\ref{gt}),
as such parameters have similar role in the formalism.

\section{Conservative system}
It is useful to describe shortly the solitons and their stability
in the conservative case ($\gamma_2 = 0$).
One dimensional conservative case has been considered by using a 
variational approach (VA) in \cite{SM}.  Using an exact approach, the 2D case with 1D nonlinear optical
lattice was studied in \cite{Fibich1},  where the authors have
considered the case with attractive background nonlinearity
($\gamma_1 > 0$). Looking for perspective applications to BEC, we will
consider here the 2D problem with 1 D nonlinear optical lattice.
Following Ref.~\cite{SM}, we start our analysis using the VA formalism.

With $u(x,y,\tau) \equiv v(x,y)\exp(-{\rm i}\mu \tau)$ in (\ref{gpe1}),
and taking $\alpha_f$, $\alpha$ and $\gamma_2$ equal zero, we obtain
\begin{equation}
\mu\; v =
-\frac{\partial^2{v}}{\partial x^2} - \frac{\partial^2{v}}{\partial y^2}
- \left(\tilde{\gamma}_{0} + \gamma_1\cos(\kappa x)\right) v^3  + \omega^2y^2 v,
\label{veq}
\end{equation}
 where  we are redefining $\gamma_0$ to $\tilde{\gamma}_0\equiv \gamma_0 + \gamma_1$.
In view of our definitions in (\ref{transf1}), this implies that for attractive condensed
systems we have $\tilde\gamma_0 = \gamma_1+1/2$; and, for repulsive ones,
$\tilde\gamma_0 = \gamma_1-1/2$. The sign of $\tilde\gamma_0$ gives the sign of the
background field. But, we should note that we can have situations where the same
$\tilde\gamma_0\ge 0$ can refer to attractive or repulsive condensates.
Example: \newline
$\tilde\gamma_0 = 1$ with $\gamma_1 = 1/2$ and $\gamma_0 = 1/2$ (attractive);\newline
$\tilde\gamma_0 = 1$ with $\gamma_1 = 3/2$ and $\gamma_0 = -1/2$ (repulsive).\newline
These two situations differ in (\ref{veq}), because the strength
of the oscillatory term is different. However, as the results are similar, we prefer
to analyze separately the cases of repulsive condensates with negative background field,
which occur when $0<\gamma_1<1/2$ ($0 > \tilde\gamma_0 \ge - 1/2$).\newline

The corresponding averaged Lagrangian $L$ is obtained from the density ${\cal L}$, as
\begin{eqnarray}
L &=&  \int_{-\infty}^\infty dx\int_{-\infty}^\infty dy \;{\cal L} \label{L1},
\end{eqnarray}
{\small \begin{eqnarray}
2\;{\cal L} = \mu\; v^2 - \left|\frac{\partial{v}}{\partial x}\right|^2 \hspace{-0.2cm}-
\left|\frac{\partial{v}}{\partial y}\right|^2\hspace{-0.2cm}+
\frac{\tilde{\gamma}_0 + \gamma_1 \cos(\kappa x)}{2}v^4 - \omega^2 y^2 v^2.
\end{eqnarray}}
Here, it is interesting to observe that a scaling given by $\kappa$ is applied to the observables obtained from the above equations, as the root-mean-square radius
in $x$ and $y$ directions, chemical potentials, frequencies and energies.
In order to see that, we can redefine all the observables using the variable
transformation, $\bar x\equiv \kappa x$ and $\bar y\equiv \kappa y$, such that we have
no $\kappa$ dependence in a new set of observables (represented with a ``bar") that are being calculated. This scaling essentially implies to consider $\kappa\equiv 1$ in all the equations. At the end, the physical observables will be given by the relations (\ref{transf1}) (with $\kappa=1$). For example, in the case of mean-square radius we 
will have
\begin{eqnarray}
\langle x_1^2\rangle=
\frac{\langle x^2\rangle}{4k^2}\;\;\;{\rm and}\;\;\;
\langle x_2^2\rangle=
\frac{\langle y^2\rangle}{4k^2}
\label{barobs2}
.\end{eqnarray}

Next, in our VA we consider the Gaussian ansatz
\begin{equation}\label{gauss1}
v(x,y) = A \exp{\left(-\frac{x^2}{2a_1^2}-\frac{y^2}{2a_2^2}\right)},
\end{equation}
which has the normalization given by $N=\pi a_1 a_2 A^2$. The 
corresponding averaged Lagrangian is given by
\begin{eqnarray}
L &=&  \int_{-\infty}^\infty dx\int_{-\infty}^\infty dy \; {\cal L}
=\frac{N}{2}\left[\mu - \left(\frac{1}{2 a_1^2} +
\frac{1}{2 a_2^2}\right)  \right.
\nonumber \\ &-&
\left.
\frac{\omega^2 a_2^2}{2} +
\frac{ N}{4\pi a_1 a_2}\left( \tilde{\gamma}_0 +
\gamma_1 e^{-\kappa ^2 a_1^2/8}\right)
\right].\label{L}
\end{eqnarray}
From the Euler-Lagrange equations for the parameters,
$\partial L/\partial N = 0$ and $\partial L/\partial a_{i=1,2} = 0$, we
obtain:
\begin{eqnarray}\label{2mu}
2\mu = \frac{1}{a_1^2} + \frac{1}{a_2^2} - \frac{N}{\pi a_1
a_2}\left( \tilde{\gamma}_0 + \gamma_1 e^{\frac{-\kappa ^2 a_1^2}8}\right) +
{\omega^2a_2^2},
\end{eqnarray}
\begin{eqnarray}\label{N}
N &=& \frac{4\pi a_2}{a_1 [\tilde{\gamma}_0 + \gamma_1
e^{-\kappa ^2a_1^2/8}(1 + \frac{\kappa ^2a_1^2}{4})]},\\
\omega^2 a_2^4 &+&
\frac{a_2^2}{a_1^2}
\frac{\left(\tilde{\gamma}_0 + \gamma_1 e^{-\kappa ^2
a_1^2/8}\right)}{[\tilde{\gamma}_0 + \gamma_1
e^{-\kappa ^2a_1^2/8}(1 + \frac{\kappa ^2a_1^2}{4})]} - 1 = 0.
\label{eqa2}\end{eqnarray}
In the case that $\omega=0$, this set of equations, for $\mu$, $N$, and
$a_2$, can be expressed in terms of $a_1$, as
{\small \begin{eqnarray}
\label{mu0}
\hspace{-1cm}\mu_0 &=&  -\frac{1}{a_1^2}\left(
\frac{
{\tilde{\gamma}_0 + \gamma_1 e^{-\kappa ^2 a_1^2/8}}\left[1
-\frac{\kappa ^2a_1^2}{8}\right]}
{
\tilde{\gamma}_0 + \gamma_1 e^{-\kappa ^2a_1^2/8}
\left[1 + \frac{\kappa ^2a_1^2}{4}\right]
}
\right)
\\
a_{2,0}&\equiv&
{a_1\sqrt{
\frac{\tilde{\gamma}_0  + \gamma_1 e^{-\kappa ^2a_1^2/8} (1+ \frac{\kappa ^2 a_1^2}{4})}
{\tilde{\gamma}_0 + \gamma_1 e^{-\kappa ^2 a_1^2/8}}
}}
\label{a20}\\
\label{N0}
N_0&=&\frac{4\pi}
{\sqrt{\left[
\tilde{\gamma}_0 + \gamma_1 e^{-\kappa ^2a_1^2/8}(1 + \frac{\kappa ^2a_1^2}{4})
\right] \left[{\tilde{\gamma}_0 + \gamma_1 e^{-\kappa ^2 a_1^2/8}}\right]}}.
\end{eqnarray}
}
For the case that $\omega\ne 0$, the relation for $a_2$ in terms of $a_1$ can be
obtained from (\ref{eqa2}) and (\ref{a20}):
\begin{eqnarray}
a_2 =
\frac{1}{\omega a_{2,0}}\sqrt{\left[\sqrt{\frac{1}{4} + {\omega^2 a_{2,0}^4}}
- \frac{1}{2}\right]}.
\label{eqa2w}
\end{eqnarray}
Equations (\ref{eqa2w}), (\ref{2mu}) and (\ref{N}) form the set of equations for
$\omega\ne 0$.

Next, we consider separately the cases of attractive systems, with ${\tilde\gamma}_0  =
\gamma_1+\frac{1}{2} >0$,
and repulsive ones with $\tilde{\gamma}_0  = \gamma_1-\frac{1}{2}<0$.

\subsubsection{\bf Attractive condensate ($\tilde{\gamma}_0=\gamma_1 +1/2 $)}

This case, which corresponds to $\gamma_0 = 1/2$ and $\gamma_1>0$, has been investigated recently in \cite{Fibich1}.
With $\omega =0$, it is applied the set of equations (\ref{mu0}), (\ref{N0}) and
(\ref{a20}). In the general case with $\omega \ne 0$, we should consider
Eqs.~(\ref{2mu}), (\ref{N}), and (\ref{eqa2w}).

Let us first verify the analytic limiting cases of the VA expressions, for $\omega =0$:
\begin{eqnarray}
{a_{2,0}}&\to& {a_1},\;\;{\rm for}\;\;a_1<< 1\;\;{\rm and}\;\;a_1>> 1 ;\nonumber\\
{\mu_{0}}&\to&-1/{a_1^2},\;\;{\rm for}\;\;a_1<< 1\;\;{\rm and}\;\;a_1>> 1 ;\nonumber\\
{N_{0}}&\to& \frac{4\pi}{\tilde\gamma_0+\gamma_1}=\frac{4\pi}{2\gamma_1+1/2},\;\;{\rm for}\;\;a_1=0;\nonumber\\
{N_{0}}&\to& \frac{4\pi}{\tilde\gamma_0}=\frac{4\pi}{\gamma_1+1/2},\;\;{\rm for}\;\;a_1\to\infty;\nonumber
\end{eqnarray}
Limiting cases of the VA expressions, for $\omega \ne 0$:
\begin{eqnarray}
{a_{2}}&\to& \left\{ \begin{array}{ll}
{a_1},\;\;{\rm for}\;\;a_1<< 1;\\
{1}/{\sqrt\omega},\;\;{\rm for}\;\;a_1>> 1
                     \end{array} \right.
                     \nonumber\\
{\mu}&\to& \left\{ \begin{array}{ll}
-1/{a_1^2},\;\;{\rm for}\;\;a_1<< 1\\
-2/{a_1^2}+\omega\to \omega, \;\;{\rm for}\;\;a_1>> 1 ;  \end{array} \right.
\nonumber\\
{N}&\to& \frac{4\pi}{\tilde\gamma_0+\gamma_1},\;\;{\rm for}\;\;a_1=0;\nonumber\\
{N}&\to& \frac{4\pi a_2}{\tilde\gamma_0 a_1}\to 0,\;\;{\rm for}\;\;a_1\to\infty.\nonumber
\end{eqnarray}

\begin{figure}[htb]
\hspace{-0.7cm}
\includegraphics[width=9cm,height=5cm
]{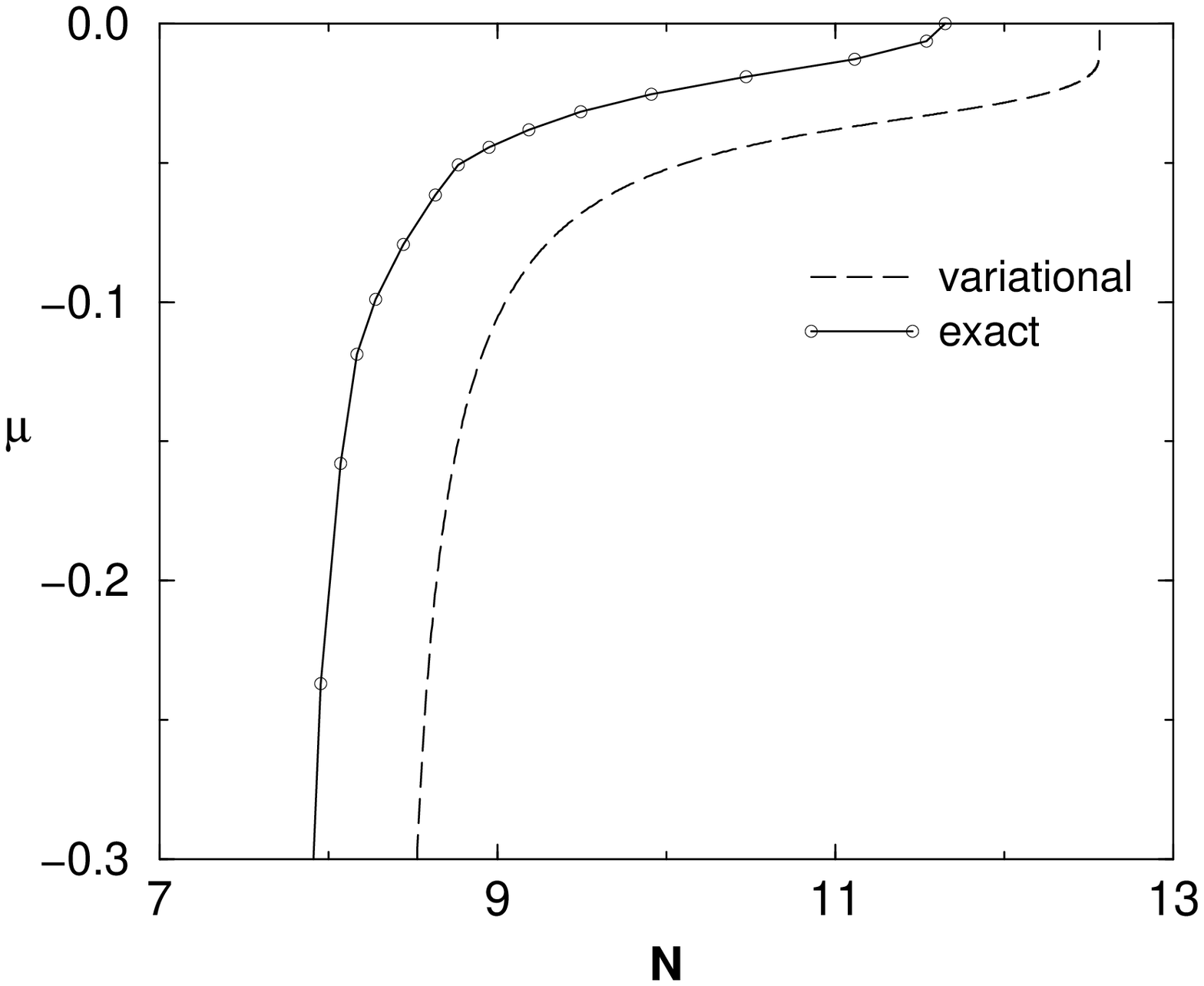}
\hspace{-0.3cm}
\includegraphics[width=8.5cm,,height=5cm
]{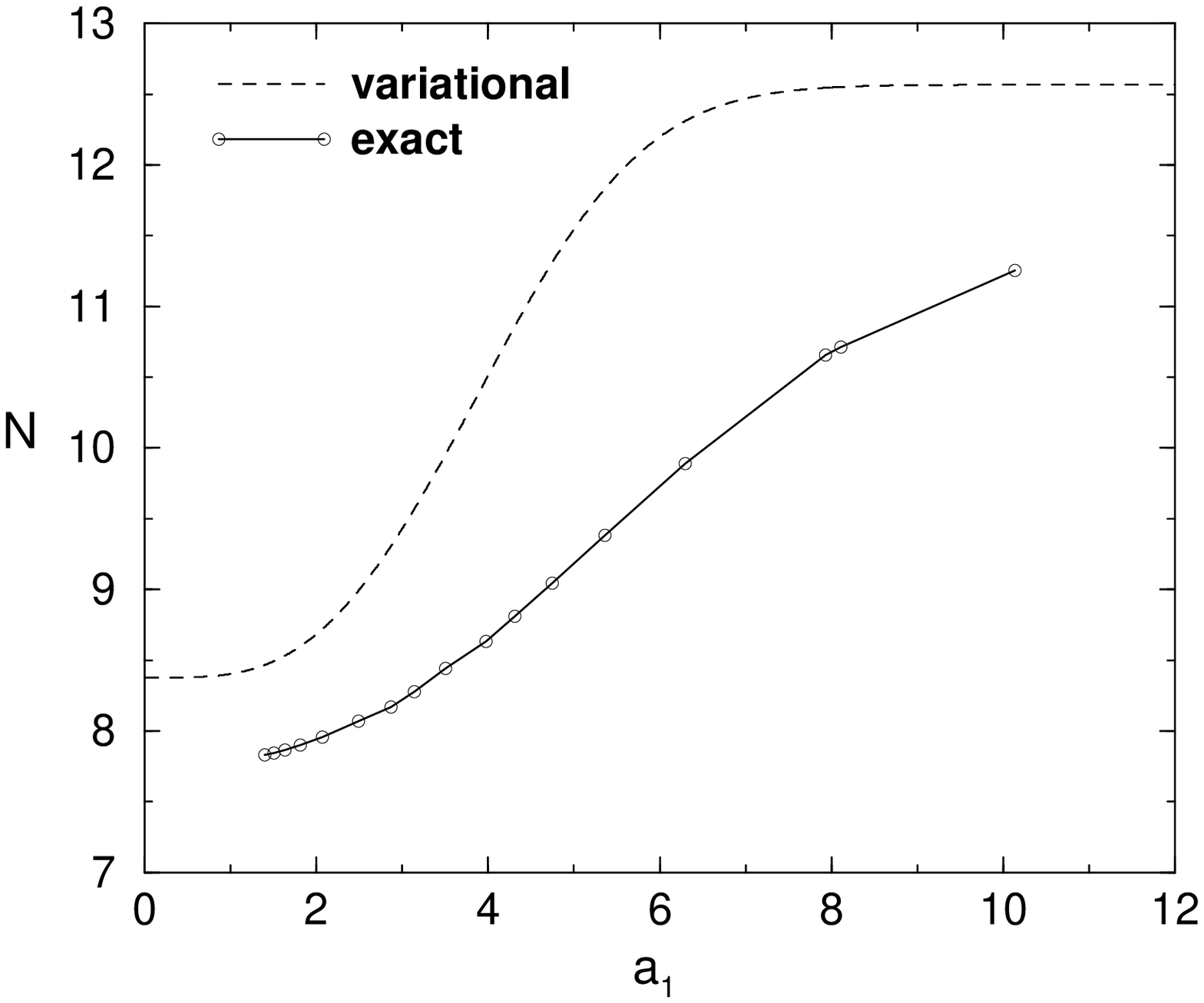}
\caption{Attractive case, with $\tilde\gamma_0=1$ and $\gamma_1=0.5$.
Results for the chemical potential $\mu$, as a function of $N$
(upper frame) and $N$ versus $a_1$ (lower frame), obtained
using variational approach (VA) and full numerical calculations.
The variational parameter for the width,
$a_1$, and the root-mean-square radius, $\sqrt{\langle x^2\rangle}$, are
related by $a_1 = \sqrt{2 \langle x^2\rangle}$
[Actually, the physical observables depend on $k$ as given by (\ref{transf1})
and (\ref{barobs2}].
} \label{fig1}
\end{figure}

\begin{figure}[htb]
\hspace{-0.7cm}
\includegraphics[width=8cm,height=5cm
]{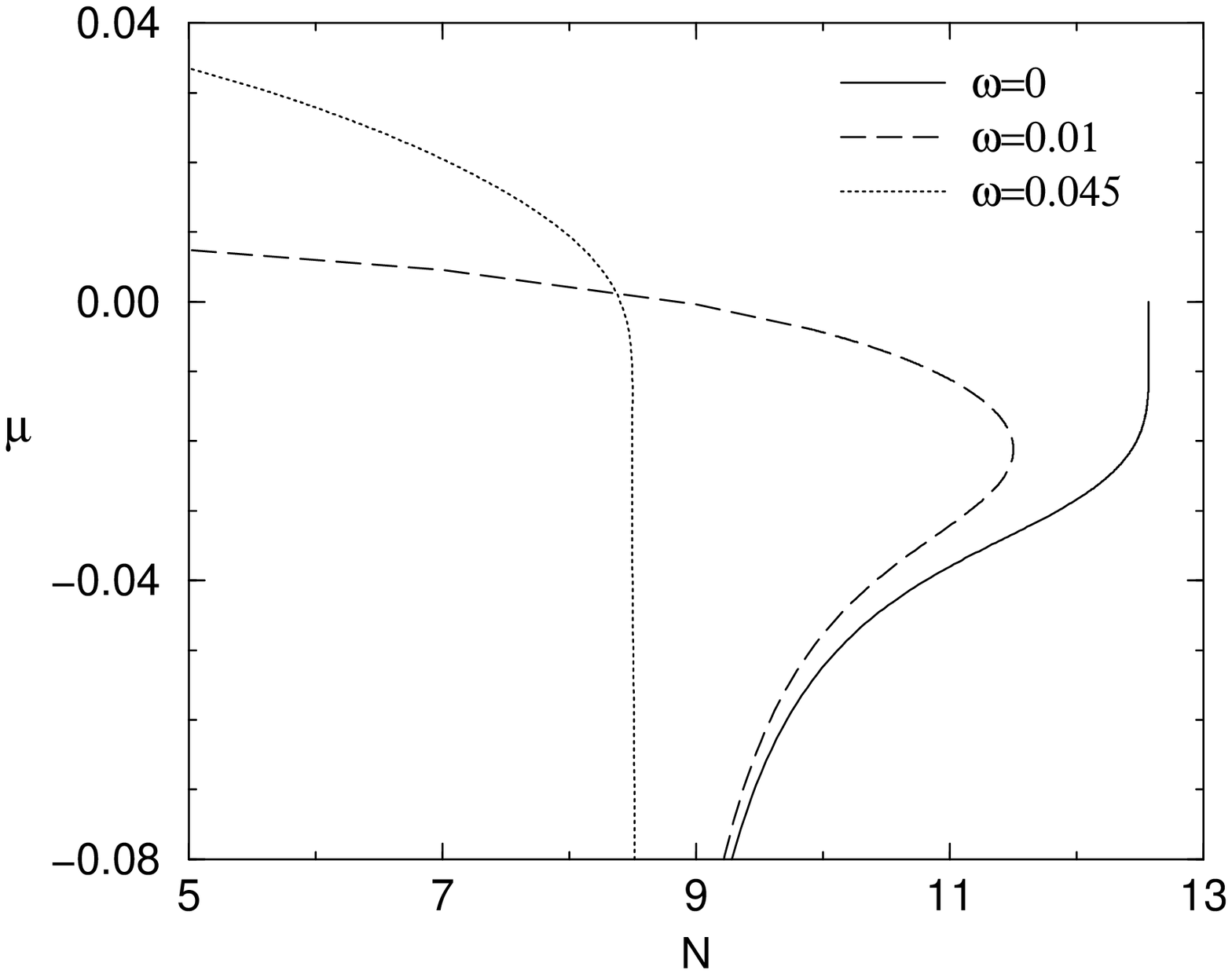}
\includegraphics[width=8cm,height=5cm
]{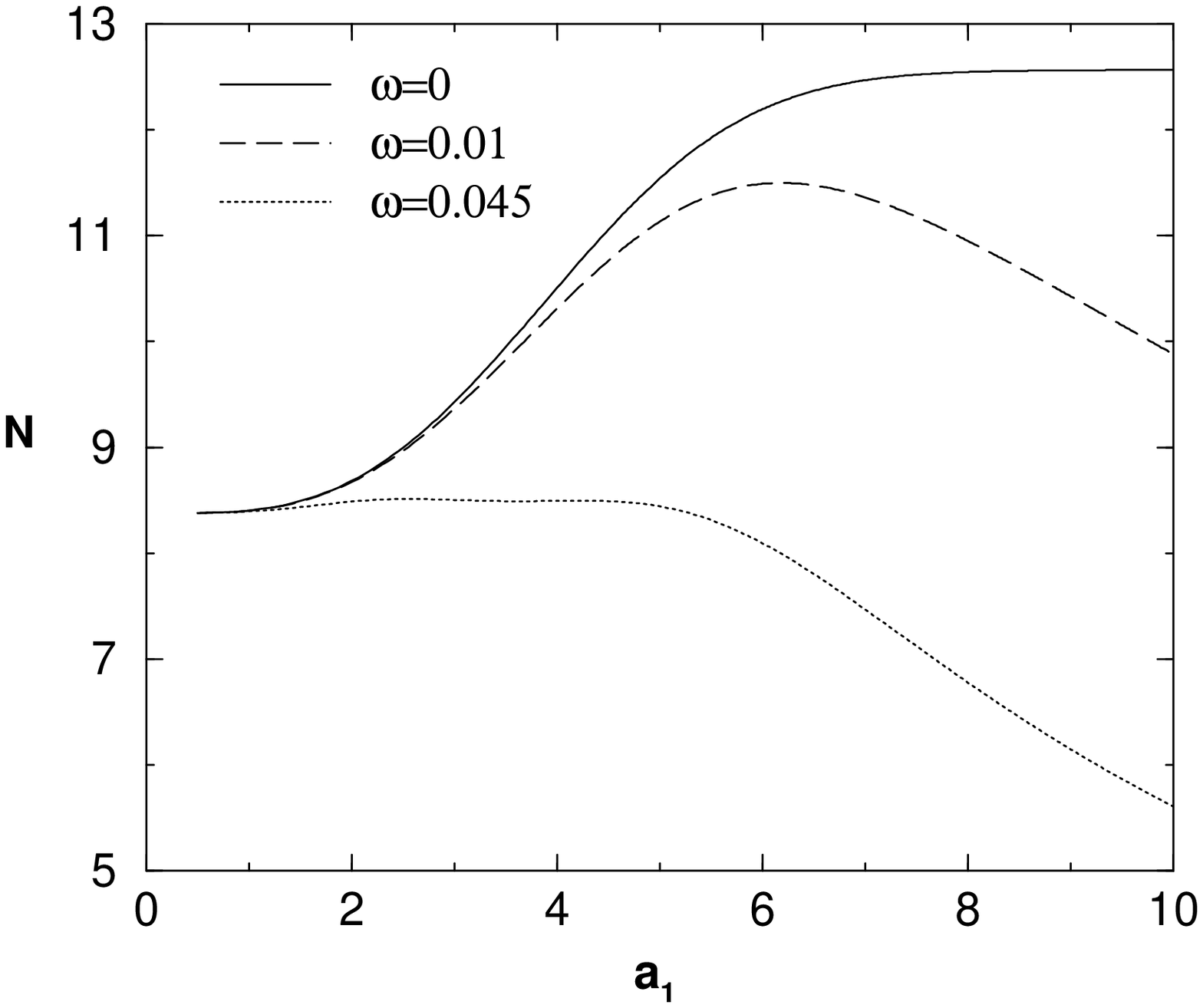}
\caption{Attractive case, with $\tilde\gamma_0=1$ and $\gamma_1=0.5$.
VA results for the chemical potential $\mu$, as a function of $N$
(upper frames) and $N$ versus $a_1$ (lower frame). The results are given
for $\mu$ near zero, considering three values of the frequency:
$\omega=$ 0 (solid line), 0.01 (dashed line) and 0.045 (dotted line).
} \label{fig2}
\end{figure}

In Fig.~\ref{fig1}, we plot the corresponding results for the
chemical potential $\mu$ as a function of $N$ (upper frame) and
$N$ as a function of $a_1$ (lower frame). Numerical soutions to
PDE results were done with algorithm presented in Ref.~\cite{Brtka}.
Considering the Vakhitov-Kolokolov (VK) criterion~\cite{VK} for the soliton stability,
$d\mu/dN < 0$, from the results given in the upper frame of Fig.~\ref{fig1}, we note
that the solitons are unstable.
This result is in agreement with  the prediction of Ref.~\cite{Fibich1}.
We note, from the VA results, that in the limit of large $a_1$ the system has a tendency
to stabilize, indicating that with just a small trapping potential we can produce a
stable region.  This behavior is shown by the VA results given in Fig.~\ref{fig2}.
The variational approach, besides an expected small quantitative shift, provides a good
qualitative picture of the results when compared with full numerical predictions. If one
is first concerned with the stability of the system (instead of the quantitative results
of the observables), the VA provides a nice and reliable picture.

In our VA, when we keep $\omega$ fixed (zero or nonzero) and increase the value of
$\gamma_1$, we observe that the general picture in respect to stability of the system
does not change. This lead us to conclude that we cannot improve the stability of the
system by increasing the strength of the lattice periodicity for attractive condensates.
In the following, we are going to analyze the cases with $\tilde\gamma_0<0$.

\subsubsection{\bf Repulsive condensate, with $\tilde\gamma_0<0$ ($\tilde\gamma_0 = \gamma_1-1/2$).}

We should remind that by repulsive condensate we mean an
atomic system where the particles have originally positive two-body scattering length,
such that in (\ref{g}) we have $g_0<0$; or $\gamma_0=-1/2$. So, given $\gamma_1$ (parameter of the spatial periodic variation of the atomic scattering length),
$\tilde\gamma_0 = \gamma_1 - 1/2$. And, if we also consider a negative background, such that $\tilde\gamma_0<0$, $\gamma_1$ will be restricted to the interval $0<\gamma_1<1/2$.

Some other limitations are applied in the parameters, considering that the widths and $N$
must be real positive quantities.  The relation between the widths $a_2$ and $a_1$, Eq.~(\ref{a20}) for $\omega=0$,  implies in a limitation to the values of $a_1$:
\begin{equation}
 e^{-\frac{\kappa ^2 a_1^2}{8}} \ge \frac{1}{2\gamma_1}-1,\;\;\to\;\; a^2_{1,max}=
 {\frac{8}{\kappa^2}
\ln\left(\frac{\gamma_1}{\frac12-\gamma_1}\right)}
\label{repcond}.\end{equation}
This limit, $a_{1,max}$, is necessary in order to have  $a_2$ and $N$ real and positive quantities for any values of $\omega$.  It will also restrict the actual values of the parameter $\gamma_1$ to
1/4$<\gamma_1<$1/2. 
The cases with $\gamma_1 > 1/2$ are also allowed, in principle, without upper limit for $a_1$. 
However, such cases will correspond to positive background field, $\tilde\gamma_0>0$, that have 
already been considered in the previous subsection.

In view of the above, let us also verify in this case the analytic VA limits.
For $\omega =0$:
\begin{eqnarray}
{a_{2,0}}&\to&
\left\{ \begin{array}{ll}
{a_1},\;\;&{\rm for}\;\;a_1<< 1; \\
\infty,\;\;&{\rm for}\;\;a_1=a_{1,max}.\end{array}\right.
\nonumber\\
{\mu_{0}}&\to&\left\{ \begin{array}{ll}
-1/{a_1^2},\;\;&{\rm for}\;\;a_1<< 1;\\
1/{(2a_1^2)},\;\;&{\rm for}\;\;a_1=a_{1,max}.
\end{array}\right.\nonumber\\
{N_{0}}&\to& \left\{ \begin{array}{ll}
{8\pi}/{(4\gamma_1-1)},
\;\;&{\rm for}\;\;a_1=0;\\
\infty,\;\;&{\rm for}\;\;a_1=a_{1,max}.\end{array}\right.
\nonumber
\end{eqnarray}
And for $\omega \ne 0$:
\begin{eqnarray}
{a_{2}}&\to& \left\{ \begin{array}{ll}
{a_1},\;\;&{\rm for}\;\;a_1<< 1;\\
{1}/{\sqrt\omega},\;\;&{\rm for}\;\;a_1=a_{1,max}.
                     \end{array} \right.
\nonumber\\
{\mu}&\to& \left\{ \begin{array}{ll}
-1/{a_1^2},\;\;&{\rm for}\;\;a_1<< 1 ;\\
1/{(2a_1^2)}+\omega, \;\;&{\rm for}\;\;a_1= a_{1,max}.
\end{array} \right.
\nonumber\\
{N}&\to& \left\{ \begin{array}{ll}
{8\pi}/{(4\gamma_1-1)},\;\;&{\rm for}\;\;a_1=0 ;\\
{32\pi}/{[(1-2\gamma_1)\sqrt{\omega}\kappa^2 a^3_{1,max}}],
\;\;&{\rm for}\;\;a_1=a_{1,max}.
\end{array}\right.\nonumber
\end{eqnarray}

\begin{figure}[htb]
\includegraphics[width=8cm,height=5cm
]{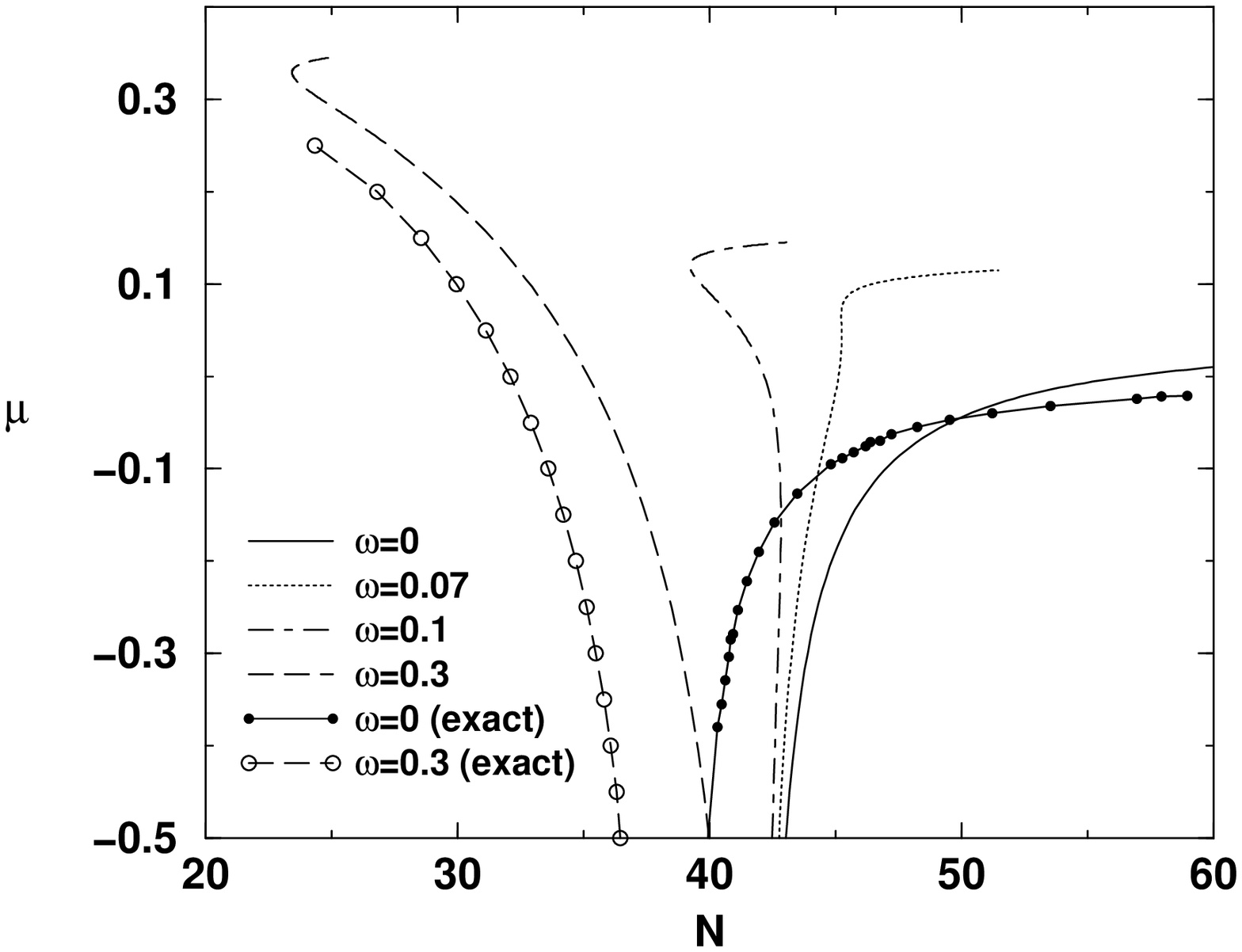}
\includegraphics[width=7.5cm,height=4.5cm
]{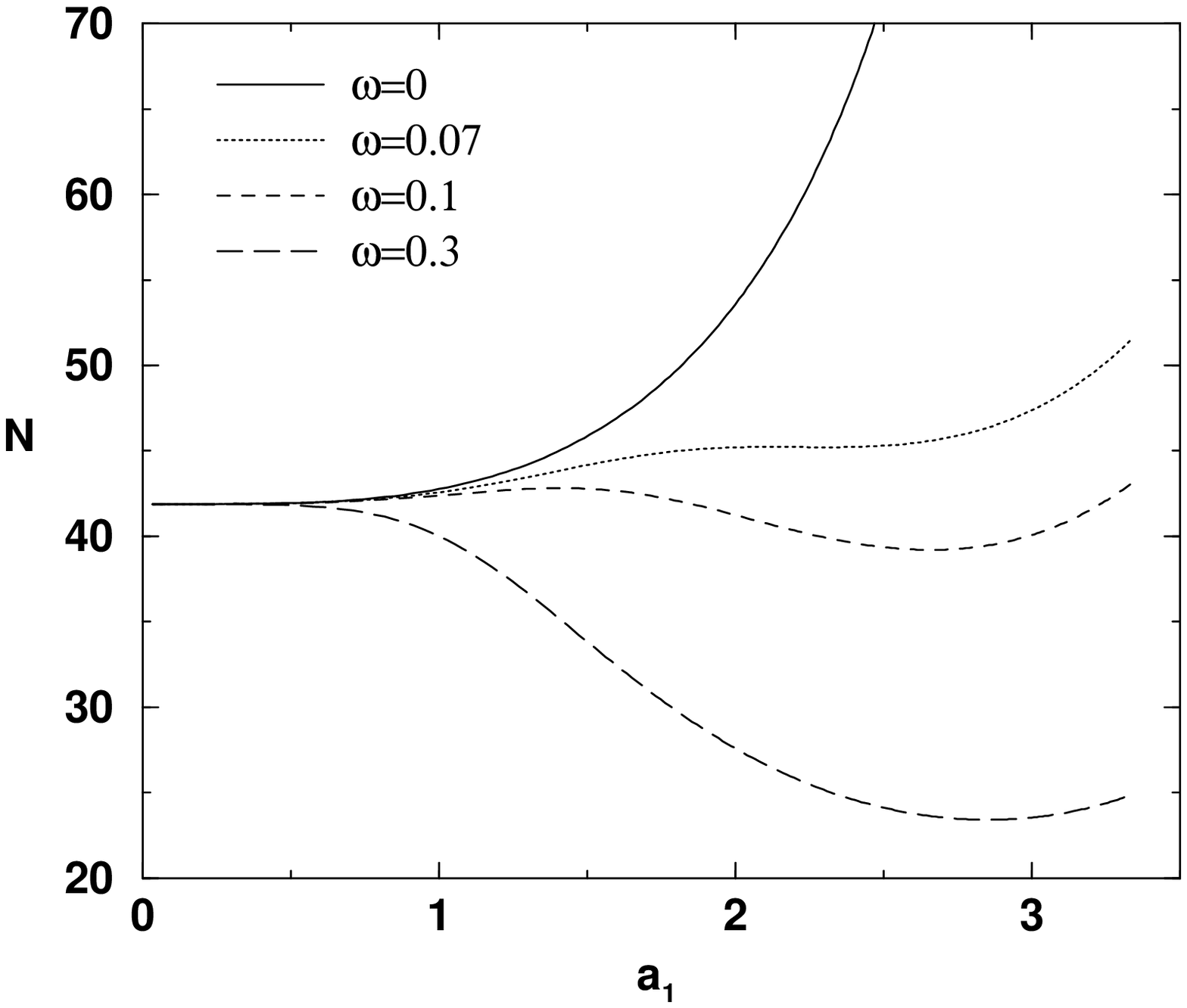}
\caption{Repulsive case, with $\tilde\gamma_0=-0.1$ and $\gamma_1=0.4$, for $\mu$ versus $N$ (upper frame), and
$N$ versus $a_1$ (lower frame).
In both frames, we show the results using the variational
approach, for $\omega=$ 0, 0.07, 0.1 and 0.3.
In the upper frame, the exact PDE results are also shown in
two cases: $\omega=0$ (for which the system is unstable) and $\omega=0.3$ (for which the system is stable). In this last case,
near the region where the VA presents a small unstable
branch ($22<N<25$), our exact numerical results are shown
only for $N>24$. As observed, the VA is giving a
general picture of the exact solutions.
} \label{fig3}
\vspace{-0.5cm}
\end{figure}

In Fig.~\ref{fig3}, we plot $N$ versus $a_1$ and the chemical potential
$\mu $ versus $N$, for $\tilde\gamma_0=-0.1$ and $\gamma_1=0.4$,
considering VA and four values of $\omega$ (0, 0.07, 0.1, 0.3).
In the case of $\omega = 0$, we also include results obtained from exact
PDE calculations.
Following the VK criterion for stability, $d\mu/dN < 0$, we notice that
stable regions start to appear with $\omega\approx 0.1$. With
$\omega> 0.3 \kappa^2$ the unstable regions almost disappear.
However, as one can observe in the lower frame, the width $a_1$ is quite
limited due to the condition (\ref{repcond}).
The observables $\mu$ and $a_i$ depend on the wave parameter $k$ of
to the spatial periodic variation of the atomic scattering length through
the scaling relations (\ref{transf1}) and (\ref{transf2}) with $\kappa=1$.
However, contrary to some discussions and conclusions of Ref.~\cite{Fibich1},
specific values of the parameter $k$ cannot affect the conclusions on
stability. In such cases of conservative systems, the stability results
from combined effects given by the parameters $\tilde\gamma_0$, $\gamma_1$
and $\omega$.
Our main conclusion is that, without the trapping potential (included in the
$y-$direction), taking $\omega=0$, the optical lattice cannot
stabilize the solutions, neither for repulsive nor for attractive condensates.

\begin{figure}[htb]
\includegraphics[width=8cm,height=5cm
]{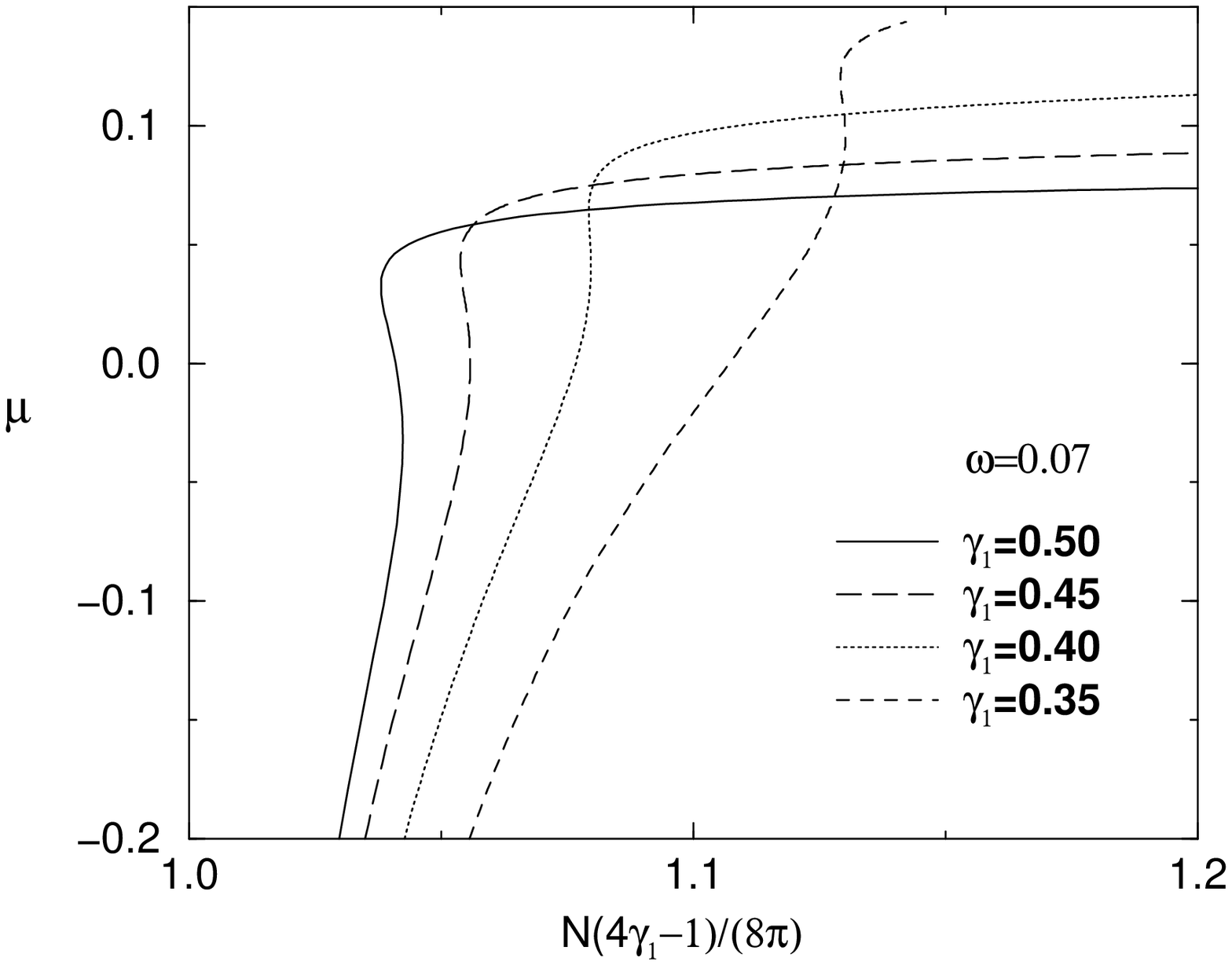}
\includegraphics[width=8cm,height=5cm
]{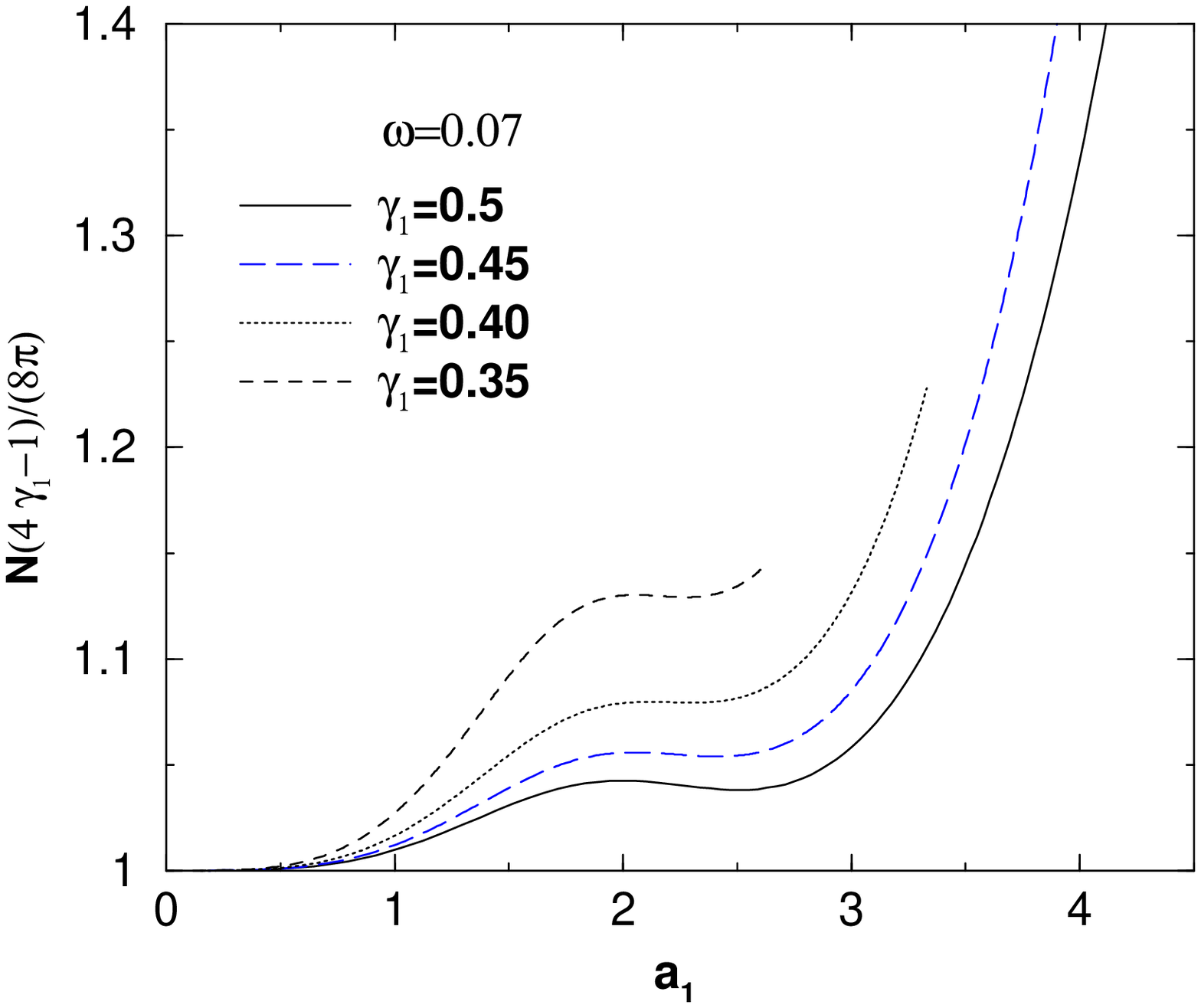}
\caption{VA results for the repulsive case, with $\tilde\gamma_0=-0.1$
and $\omega$ fixed to 0.07, considering $\gamma_1=$ 0.35, 0.4, 0.45 and 0.5.
In the upper frame we have $\mu$ versus $N/N(a_1=0)$; and, in the lower frame,
$N/N(a_1=0)$ versus $a_1$.
} \label{fig4}
\end{figure}

\begin{figure}[htb]
\includegraphics[width=8cm,height=5cm]{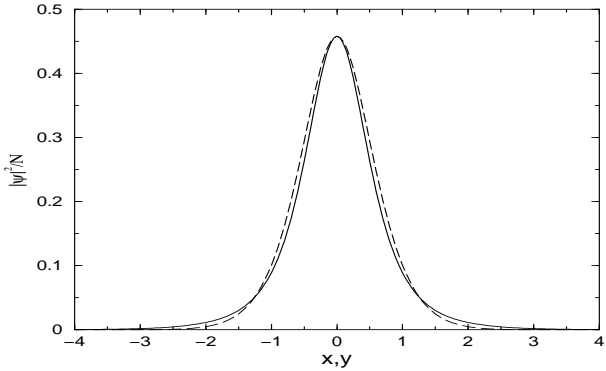}
\caption{The soliton profile in the stable region predicted by the
VK criterion for $\omega =0.3$, $\tilde{\gamma}_0 = -0.1$,
$\gamma_1 = 0.4$, $\mu =0.1$, $N \approx 33$ at the time $t=50$.
Solid line is x-direction, dashed line is y-direction. }
\label{fig5}
\end{figure}

In order to further check the role of the optical lattice, for the repulsive case
we also investigate the case with constant $\omega$ and different values of $\gamma_1$.
From the results shown in Fig.~\ref{fig3}, for $\gamma_1=0.4$, we found appropriate to
consider $\omega=0.07$, which has a marginal stability near $\mu\approx 0.05$.
The results are shown in Fig.~\ref{fig4}, where we first observe that a larger $\gamma_1$ can help to allow the width $a_1$ to increase, within the limiting condition (\ref{repcond}). However, the marginal stability remains for corresponding
different values of the chemical potential. In order to keep the plots of
Fig.\ref{fig4} for different values of $\gamma_1$ in the same frames, we have normalize the number $N$ such that it is equal to one when $a_1$ is zero.

The plots of the evolution of profiles are presented for $\omega
=$ 0 and 0.3 in Fig.~\ref{fig5}, confirming the VK prediction. The
results using the VA have good agreement with PDE prediction, as
shown for $\omega =$0 and 0.3. The profile at $t=50$ is
practically undistinguishable from the initial soliton form.

\section{Evolution of 2D soliton under 1D periodic nonlinearity and dissipation}
In this section, we will consider the case we have $\gamma_2\ne 0$ in (\ref{gpe1}).
To study the dynamics of a 2D soliton with 1D periodic nonlinearity and dissipation, we also apply a variational approach and full numerical calculations.
In the Gaussian ansatz (\ref{gauss1}) we should also include parameters related to
dissipative effects and initial conditions. For a bright soliton, the ansatz can be taken
in the following form:
\begin{eqnarray}\label{gauss2}
u &=& A \exp\left[-\frac{(x-x_{0})^2}{2a_1^2} - \frac{y^2}{2a_2^2}\right]\times\nonumber\\
 &\times&
\exp\left[{\rm i}b_1(x-x_{0})^2 + {\rm i}b_2 y^2 + {\rm i}\phi(\tau)
\right],
\end{eqnarray}
where $b_i$ ($i=$1,2) are related to dissipative effects, with $x_0$ and the phase $\phi$
related to initial condition.

The Lagrangian density for Eq.(\ref{gpe1}) is
\begin{eqnarray}
{\cal L}(x,\tau) = \frac{\rm i}{2}(u_{\tau}u^{\ast} -
u_{\tau}^{\ast}u) - |u_{x}|^{2} - |u_{y}|^{2}+
\frac{\gamma(x)}{2}|u|^4,
\end{eqnarray}
where $\gamma(x)$ is given by (\ref{gamma}). Next, from the ansatz
(\ref{gauss2}), we obtain the corresponding averaged Lagrangian:
\begin{eqnarray}
L &=& \int_{-\infty}^{\infty}dx\int_{-\infty}^{\infty}dy {\cal L} \nonumber\\
&=& - \frac{\pi}{2} A^2 a_1 a_2
\left[a_1^2(b_{1\tau} + 4 b_1^2) + \frac{1}{a_1^2}
      + a_2^2(b_{2\tau} + 4 b_2^2)
\right. \nonumber\\ &+&\left.
\frac{1}{a_2^2} + 2\phi_{\tau} -
\frac{A^2}{2}\left(\tilde{\gamma}_0 +
\gamma_1 \cos(\kappa x_0)e^{-\kappa ^2 a_1^2/8}\right)\right]
\label{lagr2}.
\end{eqnarray}
The equations for the soliton parameters $\eta_{i} = [A,a,b,\phi]$
in the VA are derived from(see for example\cite{FAGT})
\begin{equation}
\frac{\partial L}{\partial\eta_{i}}- \frac{d}{d\tau}\frac{\partial
L}{\partial\eta_{i,\tau}} =
\int_{-\infty}^{\infty}dx\int_{-\infty}^{\infty}dy
[R\frac{\partial
 u^{\ast}}{\partial\eta_{i}} + c.c],
\end{equation}
where the perturbation term $R$ is
\begin{equation}
R = -{\rm i}\gamma_2 (1+\cos(\kappa x))|u|^2 u + {\rm i}\alpha_f u.
\end{equation}
Here we are taking into account a linear amplification term ($\alpha_f$) describing
the atoms feeding. Finally, from the above, we obtain the following system of five
coupled ordinary differential equations (ODE) to be solved for the parameters of
our variational approach (VA):
\begin{widetext}
\begin{eqnarray}
(A^2 a_1 a_2)_{\tau}&=&-\gamma_2 A^4 a_1 a_2  e^{-\kappa^2a^2/8}\cos(\kappa x_0)
- \gamma_2 A^4 a_1 a_2 + 2\alpha_f A^2 a_1 a_2 ,\\
(A^2 a_1^3 a_2 )_{\tau} &=& 8A^2 a_1^3 a_2 b_1 -
\frac{\gamma_2}{2} A^4 a_1^3 a_2[1  + \frac{1}{4}
\cos(\kappa x_0)(4 - \kappa ^2 a_1^2)e^{-\kappa^2a_1^2/8}] + 2\alpha_f A^2 a_1^3 a_2,\\
(A^2 a_1 a_2^3 )_{\tau} &=& 8A^2 a_1 a_2^3 b_2 -
\frac{\gamma_2}{2}A^4 a_2^3 a_1[1 + e^{-\kappa^2 a_1^2/8}\cos(\kappa x_0)] +  2\alpha_f A^2 a_1 a_2^3,\\
b_{1\tau} &=& \frac{1}{a_1^4} -4b_1^2 - \frac{\tilde{\gamma}_0
A^2}{4a_1^2} - \frac{\gamma_1 A^2}{4a_1^2}\cos(\kappa x_0)
e^{-\kappa^2a_1^2/8}(1 + \frac{\kappa^2 a_1^2}{4}) ,\\
b_{2\tau} &=& \frac{1}{a_2^4} -4b_2^2 - \frac{\tilde{\gamma}_0
A^2}{4a_2^2} - \frac{\gamma_1 A^2}{4a_2^2}\cos(\kappa x_0)
e^{-\kappa^2 a_1^2/8}.
\end{eqnarray}
By taking into account that the norm $N=\pi A^2 a_1 a_2$,
with $i,j =$ 1,2 ($i\ne j$), we have:
\begin{eqnarray}
N_{\tau} &=& -\frac{\gamma_2 N^2}{\pi a_1 a_2}\left[1 + e^{-\kappa^2
a_1^2/8}\cos(\kappa x_0)\right] + 2\alpha_f N,\\
(a_i^2)_{\tau} &=& 8a_i^2 b_i  + \frac{\gamma_2 Na_i}{2\pi
a_j}\left[1 +  e^{-\kappa^2 a_1^2/8}\cos(\kappa x_0)
\left(1+\delta_{i,1}\frac{\kappa^2a_1^2}{4}\right)\right]\\
b_{i,\tau} &=& \frac{1}{a_i^4} - 4 b_i^2 - \frac{N}{4\pi
a_i^3a_j}\left(\gamma_0 +  \gamma_1\left[1 + e^{-\kappa^2
a_1^2/8}\cos(\kappa x_0)
\left(1+\delta_{i,1}\frac{\kappa^2a_1^2}{4}\right)\right]\right)
\end{eqnarray}
\end{widetext}

In the next, we present some of our results, when considering periodic
nonlinearity with dissipative effects.
 Considering the scaling of observables with $\kappa$, discussed 
for the conservative systems in section III, which
can also be verified in the present case, we have the corresponding
transformation $b_i\to b_i/\kappa^2$.

\begin{figure}[htb]
\includegraphics[width=8cm,height=5cm]{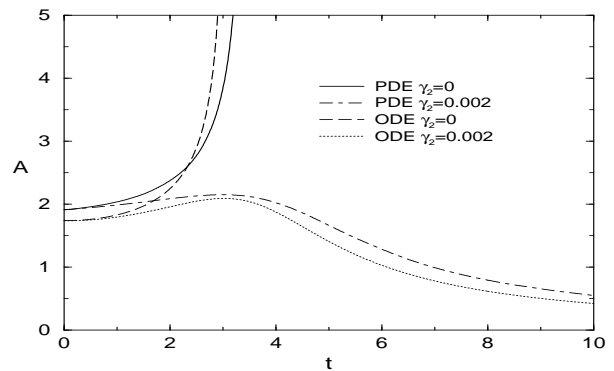}
\caption{Results for the amplitude as a function of $\tau$, using
variational approach (VA) and full numerical calculations. The VA
and the full numerical calculations have the value of $\mu$ fixed
to the same value for $\tau=0$, implying in a small shift of
$A(0)$, as shown by the results.} \label{fig6}
\end{figure}

In Fig.~\ref{fig6}, we have results for the of full numerical simulations
(PDE) for the evolution of  the matter wave packet under combination of the
conservative and dissipative nonlinear optical lattice in the case
of the attractive condensate $\gamma_0 = 1/2$. As we can see the
collapse is arrested by the dissipative nonlinear optical lattice.
The results are compared with the prediction of the VA approach (ODE).
We observe a good agreement of the VA with full-numerical calculations.

\begin{figure}[htb]
\includegraphics[width=8cm,height=5cm]{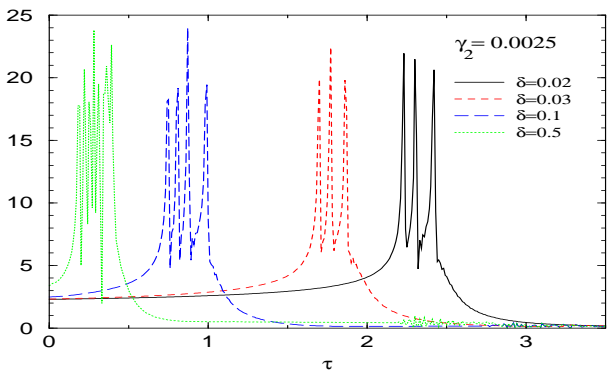}
\caption{Results for the amplitude as a function of $\tau$,
showing the collapsing behavior for $\kappa =4\pi$. The dissipation is
fixed to $\gamma_2=0.0025$ and $\delta$ is increasing from 0.02 to
0.5.
} \label{fig7}
\end{figure}

We also have investigated the role of a deviation $\delta$ of the
given norm from the critical norm, in the initial wave packet:
$A\to A(1+\delta)$.
The results of the full numerical simulations
are presented in Fig.~\ref{fig7}. Increasing the deviation $\delta$ from
the critical norm, multiple peaks are observed, corresponding to revivals
of the wave packet during the collapse. The number of peaks growths as
$\delta$ varies from 0.02 to 0.5.
The focusing-defocusing cycles connect the action of the
periodically varying in the space with the inelastic three-body
interactions.
In linear conservative optical lattice, with inelastic three-body interactions,
the focusing-defocusing oscillations have been studied in Ref.~\cite{AS1}.

\begin{figure}[htb]\begin{center}
\includegraphics[width=7cm,height=5.5cm]{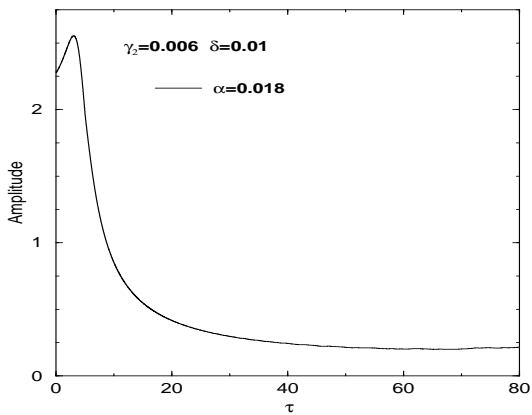}
\end{center}
\caption{Soliton stabilization via compression effect,
with $\tau_c=5$ and $\alpha=0.018$ in Eq.~(\ref{gt}).
The other parameter are $\gamma_2=0.006$, $\delta=0.01$, and
$\kappa =4\pi$. 
}
\label{fig8}
\end{figure}

The spreading out of the pulse after the collapse is arrested,
observed in Fig.~\ref{fig6}, can be compensated, as we note previously,
by an adiabatic variation of the background scattering length, described
by a time variation of $\gamma_0$. When we consider a time
dependent $\gamma_0$, as given in Eq.~(\ref{gt}), the feeding term
parameter $\alpha_f$ should be zero, because one can show
(with a redefinition of the wave-function) that it has a similar effect.
In Fig.~\ref{fig8}, we show our full numerical results confirming the
stabilization of the condensate after the collapse was arrested. The
mechanism of this stabilization was given by an appropriate tunning
of the parameters $\alpha$ and $\tau_c$ of Eq.~(\ref{gt}).

\section{Conclusion}

Dynamics and  stability of matter-wave solitons in the mixture of
conservative and dissipative nonlinear optical lattices are
investigated, considering 2D BEC, with 1D conservative plus dissipative
nonlinear optical lattices.

In the first part of this work, it was analyzed conservative systems,
with nonlinear optical lattices, for attractive and repulsive condensates.
It was clarified the role of the scales when calculating the observables
as the chemical potential and the widths. Our conclusion is that, in a
2D system, a nonlinear periodic lattice in one direction by itself cannot give
stable solutions, satisfying the VK criterium~\cite{VK}.  Such periodic
lattice in the $x-$direction cannot compensate the collapsing effect
which results from the other dimension. We verify that stable solutions 
can be obtained by controlling the soliton with a harmonic trap in the $y-$direction.
For repulsive condensates, the 2D stable soliton can exist in the
geometry with 1D nonlinear optical lattices in one direction and harmonic
trap in the other direction.

In the second part of the work we analyze the dynamics of the above 2D
system, with periodic nonlinearity in $x-$direction and without trap
in the $y-$direction, when we add non-conservative nonlinear optical lattice
terms.
We show that the collapse of the condensate can be arrested by a
dissipative periodic nonlinearity.
To study the evolution of 2D wavepacket we apply the time-dependent variational 
approach. To compensate the wavepacket broadening,
the adiabatic time variation of scattering length is used. It is shown that  the
metastable dissipative soliton can exist in 2D condensate with 1D
periodic nonlinearity. Analytical predictions are confirmed by the
numerical simulations of full 2D GP equation.

\section*{Ackhowledgments}
We thank Funda\c c\~ao de Amparo \`a Pesquisa do Estado de S\~ao
Paulo for partial support. LT and AG also thank Conselho Nacional
de Desenvolvimento Cient\'\i fico e Tecnol\'ogico for partial
support.

\end{document}